\documentclass{article}
\usepackage[utf8]{inputenc}

\pdfoutput=1

\usepackage[scaled]{helvet}
 
\usepackage[T1]{fontenc}
\usepackage{dirtytalk}
\usepackage{graphicx}
\usepackage{authblk}
\usepackage{amsmath}
\usepackage{hyperref}
\DeclareMathOperator*{\argmax}{arg\,max}

\usepackage[a4paper, total={6.5in, 7.5in}]{geometry}

\title{Social media usage reveals how regions recover after natural disaster}

\author[a]{Robert Eyre}
\author[b]{Flavia de Luca} 
\author[a,*]{Filippo Simini}

\affil[a]{University of Bristol, Department of Engineering Mathematics, Bristol, BS8 1UB, UK}
\affil[b]{University of Bristol, Department of Civil Engineering, Bristol, BS8 1UB, UK}
\affil[*]{To whom correspondence should be addressed; E-mail: f.simini@bristol.ac.uk}

\date{}

\begin{document}

\maketitle

\abstract{The challenge of nowcasting and forecasting the effect of natural disasters (e.g. earthquakes, floods, hurricanes) on assets,
people and society is of primary importance for assessing the ability of such systems to recover from extreme events. Traditional
disaster recovery estimates, such as surveys and interviews, are usually costly, time consuming and do not scale. Here we
present a methodology to indirectly estimate the post-emergency recovery status (‘downtime’) of small businesses in urban
areas looking at their online posting activity on social media. Analysing the time series of posts before and after an event, we
quantify the downtime of small businesses for three natural disasters occurred in Nepal, Puerto Rico and Mexico. A convenient
and reliable method for nowcasting the post-emergency recovery status of economic activities could help local governments
and decision makers to better target their interventions and distribute the available resources more effectively.}
\\

Natural disaster management is of interest to organisations that deliver aid, provide insurance, and operate in areas affected by disasters. The steps in natural disaster management are categorised into \say{phases}, with previous literature identifying anywhere from four to eight phases during the occurrence of a natural disaster \cite{neal1997reconsidering}. For the purpose of this article, we take the phases used by the Federal Emergency Management Agency (FEMA) - mitigation, preparedness, response and recovery~\cite{alexander2002principles}.
Of the four stages of the disaster life cycle, recovery is the least understood and investigated \cite{Olshansky_2012}. 

Factors such as deaths and building damage have been used to immediately quantify the magnitude of a disaster, but are not necessarily indicative of the longer-term recovery process. 
Multiple indicators for the recovery process of an area have been  proposed~\cite{Horney_2016}, 
with many of them focusing on economic indicators of business activity thought to capture the long-term efforts of a region to return to a normal state. 
In particular, we refer herein to the recovery time as \say{downtime} relying on the definition within the Performance-Based Engineering framework; i.e., \say{the time necessary to plan, finance, and complete repairs on facilities damaged in earthquakes or other disasters} \cite{Comerio_2006} and applying the downtime concept to small businesses. 

The definition of actual business downtime, %
i.e. the time in which a business is either closed or not running as expected, has been debated strongly in the literature. 
Chang's framework for urban disaster recovery \cite{Chang_2010} highlights three such definitions of recovery; (a) returning to the same level of activity before the disaster, (b) reaching the level of activity that the business would have been attained without the disaster happening, and (c) attaining some stable level of activity that is different from these definitions.

Studies on long-term business recovery have been made using large-scale economic indicators,  
such as the number of reported businesses in Kobe City over a ten year period following the Kobe City earthquake in 1995 \cite{Chang_2010}. 
These large-scale economic indicators are not as readily available or relevant for disasters of a more moderate scale, inspiring the use of alternative statistics as predictors for business recovery, such as changes in pedestrian foot traffic~\cite{harding2011variations}, manually reporting on when businesses closed and opened~\cite{campanella2007street} 
and changes in parking transactions~\cite{hino2019high}, 
allowing for a much smaller scale to be studied. 
However, surveys and traditional direct data sources of businesses downtime have either been too costly or too inconvenient for widespread use. 

Remote sensing 
has been shown to be vital in rapid damage detection and relief efforts after natural disasters.  For example, the  European Union's Copernicus Emergency Management Service and the Jet Propulsion Laboratory's Advanced Rapid Imaging and Analysis project 
use satellite imagery to provide assessment of regions devastated by different types of disasters.
The main application of post-disaster satellite imagery is on emergency response and large-scale damage estimation~\cite{Booth_2011}. 
Recent efforts to predict downtime using satellite imagery include counting the number of vehicles in imagery collected by Google Earth, correlating the presence of vehicles in business zones to the activity of businesses~\cite{DeLuca_2018}. 
This method requires to have multiple satellite images of the region over time and may not be reliable in regions with developed public transport links, where the presence (or lack) of vehicles may not always correlate to small businesses' level of economic activity. 
A desirable secondary source of data for downtime estimation should be more readily accessible (easy and cheap to collect), have a high spatio-temporal resolution and, in the context of natural disaster, be rapidly obtainable. 
It is for these reasons that 
studies have turned to social media as a source of indicators for damage prediction after a natural disaster.

Social media data has been shown to be effective at rapidly detecting the presence of natural disasters such as earthquakes and floods, however the underlying process of social media use during natural disaster is not completely understood \cite{Murthy_2017}. The micro-blogging service Twitter is often a source of related data, due to the nature of the platform - only allowing short 280 character maximum (often geo-tagged) posts promotes the exchange of informative data. Other social media and content sharing websites have also been shown to exhibit correlating behaviour to natural disaster events, such as the photo sharing website Flickr \cite{Preis_2013, Tkachenko_2017}.
Many current social-media methods rely on sentiment analysis to filter messages by relevancy and quantify the severity of response to an event \cite{Zou_2018, Murthy_2017, earle2012twitter, Vieweg_2010, sutton2014warning}. 
These methods offer rapid assessment of an event's infancy 
and are useful tools for understanding human behaviour during emergencies and to improve the delivery of hazard information in a region. Links between damage indicators such as assistance data from the Federal Emergency Management Agency (FEMA), and tweets have also been found~\cite{Kryvasheyeu_2016}. 
Mobile phone data has been used to obtain insights on human behaviour during emergencies. 
Bagrow et al. \cite{Bagrow_2011} show that mobile phone data can be used to identify how human behaviour changes during periods of unfamiliar conditions, in a variety of situations ranging from bombings to festivals.

Dynamic mapping of population through the use of mobile phone records~\cite{Deville_2014} 
have been used to measure population displacement after the 2015 Gorhka earthquake in Nepal, where negative inflows were recorded in Kathmandu for roughly 60 days after the event~\cite{Wilson_2016}. 
Mobile phone data was also used in the case of the 2010 Haiti Earthquake to show that movement actually became more predictable up to three months after the event in comparison to normal activity~\cite{Lu_2012}. 
One of the main limitations of mobile phone data is that it is usually not publicly available because of privacy concerns, hence obtaining mobile phone records for a specific region is not always possible. 
Most of the studies in the literature often focus on emergency management and rapid response within the short term, identifying sharp spikes in activity to determine the immediate impact of a disaster in a given region.

\begin{figure*}[h]
    \centering
    \includegraphics[width=\textwidth]{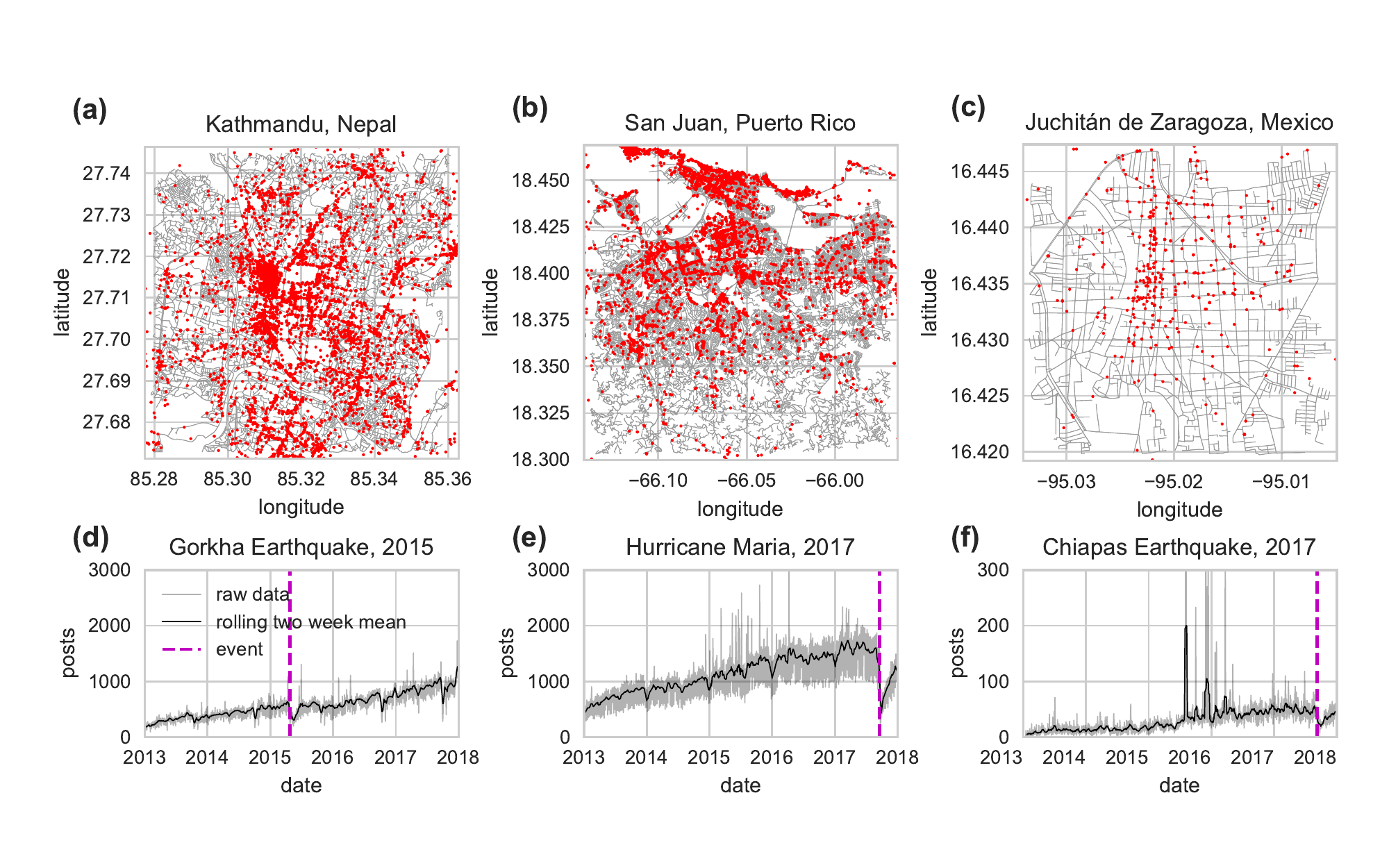}
    \caption{\textbf{(a-c)} Regions in which Facebook posts have been collected, each business highlighted in red (road networks collected from OpenStreetMap~\cite{OpenStreetMap} using OSMnx \cite{Boeing_2017}). \textbf{(d-f)} Time series for the number of posts made in each region. The vertical dashed lines denote the date of the natural disasters. }
    \label{fig:all}
\end{figure*}

In this paper we show that downtime can be estimated in real time using the public posts of local businesses on the social media site `Facebook' collected before, during and after a natural disaster, without the need for natural language processing or semantic analysis. 
In particular, we consider three natural disasters of different types (two earthquakes and one hurricane) which occurred in countries with different indicators of socioeconomic development (Nepal, Mexico and Puerto Rico). 
The locations of the businesses collected in the three regions considered are shown in Figure~\ref{fig:all}a-c and the respective time series of the total number of posts retrieved are shown in Figure~\ref{fig:all}d-f (See Materials and Methods for a description for each natural disaster)

\section{Results}

To gauge the impact of a natural disaster (or event) on the businesses within a specific region, we consider the businesses' posting activity on Facebook. 
Specifically, we compare the time series of number of posts on a business' page in the time period after the event with the typical posting activity before the event. 

\begin{figure*}[]
    \centering
    \includegraphics[width=0.5\textwidth]{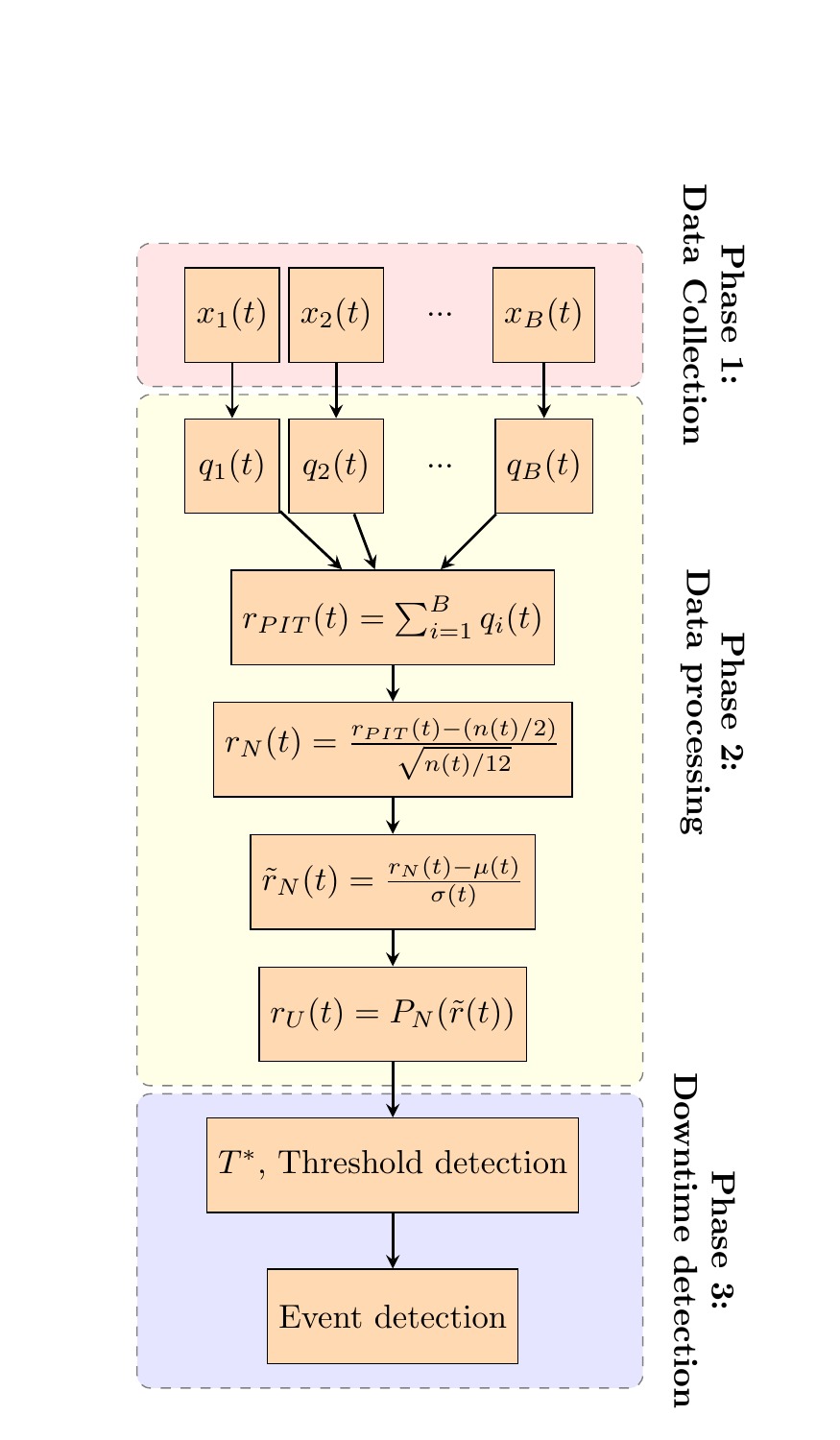}
    \caption{Schematic for the three phases of the proposed framework to detect the impact of natural disasters on small businesses.}  \label{fig:scheme}
\end{figure*}

The framework is composed of three phases
, schematically illustrated in Figure~\ref{fig:scheme}: 
(i) Data collection (Sect.~\ref{sec:datacoll}); (ii) Data processing (Sect.~\ref{sec:dataproc}); (iii) Downtime detection (Sect.~\ref{sec:downtime}). 
In phase (i) publicly available data on the businesses' posting activity on Facebook is collected. 
In phase (ii) the time series of the posting activity is processed in order to remove trends, periodicities, seasonality effects, heterogeneous posting rates and correlations, that are present in the raw data. 
In phase (iii) we define an automatic method to determine whether the posting activity presents anomalies, such as a posting rate significantly lower than the average rate, and use this to estimate the length of the downtime.

\begin{figure*}[h!]
    \centering
    \includegraphics[width=0.85\textwidth]{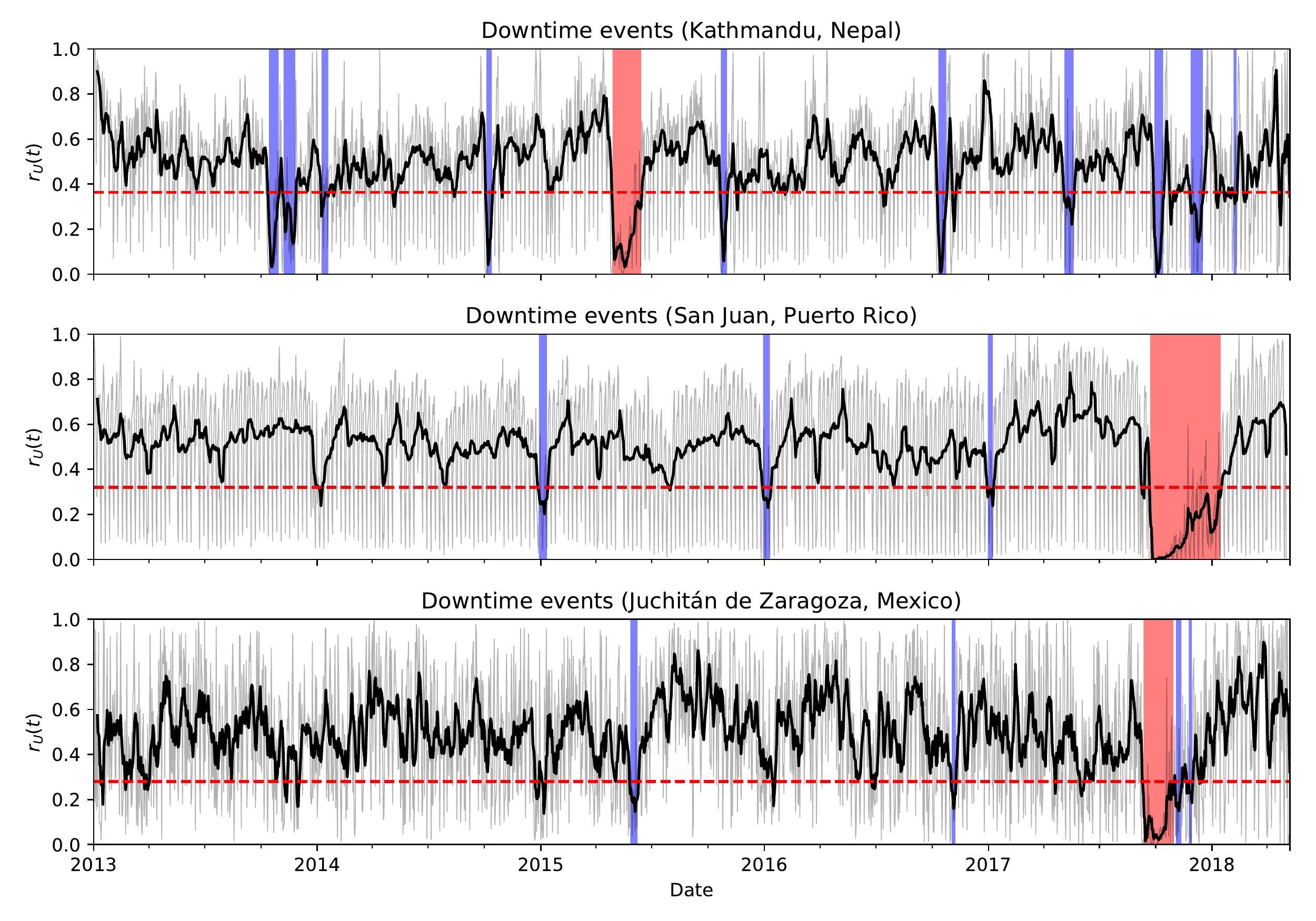}
    \caption{
    Downtime events longer than seven days detected in each region. One week rolling mean shown in black, raw data shown in grey, red dashed line denotes the threshold estimated with the `elbow method' (see Section~\ref{sec:downtime}). 
    The thresholds $T^{\ast}$ are marked with dashed horizontal red lines and the corresponding events are highlighted in red (natural disasters) and blue (other events).}
    \label{fig:final_all}
\end{figure*}

Using this framework, we find downtime in Kathmandu (50 days), Juchit\'an de Zaragoza (52 days) and San Juan (118 days) on the dates of their respective natural disaster events. 
The events detected in the three regions are shown in Figure~\ref{fig:final_all}; 
the start date, end date and the length of downtime of all the events detected are reported in 
the Supplementary Materials. 
Given that we are using a weekly rolling mean to detect the start, end and duration of the downtime, we expect our estimates to have an uncertainty of around one week. 
The following sections discuss the validation of our downtime estimates in each of the three regions considered.

\subsection{Kathmandu, Nepal} 
A downtime of 50 days is predicted over the entire city after the Gorkha Earthquake in 2015. This sits comfortably between the lower bound of 40 days found with the surveys~\cite{DeLuca_2018} 
(see Supplementary Materials), 
and the upper bound of approximately 56 days found by mobility data in \cite{Wilson_2016}. 
To further test the downtime prediction, we also applied the process to Region A (around the UNESCO Heritage site of Kathmandu Durbar Square, see Supplementary Materials) which contains the majority of the surveyed businesses, and report a downtime of 33 days - close to the 28 days of average downtime reported by the surveys in this region. 
We also detect shorter downtimes in other times of the year and this further supports the validity of our method (see Supplementary Materials). 
Indeed, the majority of the Nepalese people (84.7\%) practice Hinduism and celebrate multiple festivals throughout the year. The main festival, Dashain, is celebrated for two weeks in September or October and is of such importance to the religious and cultural identity of the country, that businesses and other organisations are completely closed for 10 to 15 days to celebrate. We detect downtime during Dashain for each year in which we have data. Additionally, we detect other festivals that follow Dashain, such as the second largest festival in Nepal called Tihar.

\subsection{San Juan, Puerto Rico} 

To determine the actual downtime in San Juan after Hurricane Maria, we took tourism data from the Puerto Rico Tourism company~\cite{puerto_tourism}, listing the monthly number of cruise passengers in the port of Old San Juan (see Supplementary Materials). 
We apply the elbow method (see Section \ref{sec:downtime}) to this data and estimate a downtime of 102 days. From the analysis of Facebook posts, the 118 days of downtime that we predict can be split into downtime from the Hurricane (103 days), and downtime from the Christmas to New Years period (14 days).
In fact, in San Juan, we found periodic downtime events during the periods between Christmas and New Year (see Supplementary Materials). Compared to Nepal, Puerto Rico has a majority Christian population, which explains the downtime during the Christmas period. 


\subsection{Juchit\'an de Zaragoza, Mexico}

The events that were found in Juchit\'an de Zaragoza 
proved to be the most difficult to verify. Whilst the final output is noisy,
we still see downtime during the Chiapas earthquake event, supported by a limited survey sent to the businesses that were collected. 
Of the businesses that responded, 
the average closure time reported (ignoring businesses that never reopened) was 63 days. This sits in between our predicted downtime of 52 days and 66 days (if we also count the 12 consecutive additional days of downtime presumed to be due to the All Saints festivity, see Supplementary Materials).

\subsection{Sensitivity analysis} 
The reported downtime for each region was calculated using all of the collected businesses that had posted at least once. 
We tested the sensitivity of our method to the overall number of businesses considered. %
We observe that the methodology gives consistent estimates when the number of businesses is large (i.e. thousands), whereas the length of downtime might be underestimated if the number of businesses is small. 
We verified this in Nepal and Puerto Rico, by randomly sampling a subset of businesses and computing the average length of downtime over 1000 realisations as a function of the sample size (see Supplementary Materials). 
Establishing a precise relationship between the number of businesses and the average downtime is difficult because that estimate depends not only on the number of the businesses but also on other factors, such as their geographic density and mean posting rate. 
Indeed, to assess to what extent the estimate of downtime depends on the posting activity of the businesses, we filtered businesses by their daily posting rate and by the total number of posts they have made. 
The results reported in the Supplementary Materials show that the overall downtime of the region is not affected by the filtering, except in cases of very high thresholds. 
In particular, we note that we can obtain an accurate estimate of the downtime with a small number of businesses (few hundreds) that post frequently (e.g. more than once per week). \\

\subsection{Downtime detection in real-time} 
 
When sufficient data from social media is available, the proposed system can be applied in real-time giving estimates of the recovery status during the weeks immediately after an event. 
We simulate the collection of data in real time by cropping our data in the weeks following the event
and we calculate $d_{\rm RT}(t)$, the real-time estimate of the downtime $t$ weeks after the event, using just the posts published until week $t$. Results for Kathmandu are shown in Figure~\ref{process_real}, for the other regions see Supplementary Materials. 
To evaluate the accuracy of the real-time estimate, we measure the root mean squared distance (RMSD) between $d_{\rm RT}(t)$ and the ground truth downtime, 
$d_{\rm GT}(t) = \min(t, d^\ast)$, where $d^{\ast}$ is the downtime estimated using all data ($d^\ast = 50$ days for Kathmandu). 
Computing the 
${\rm RMSD} = \sqrt{\sum_{t=0}^{d^\ast} (d_{\rm RT}(t) - d_{\rm GT}(t))^2 / d^\ast}$ 
from the date of the event $t = 0$ until $t = d^\ast$, i.e. for a time period of the length of the actual downtime from the date of the event, we obtain 2.75, 0.88, 3.5 days for Kathmandu, San Juan and Juchit\'an de Zaragoza respectively. 
In all cases the errors are within the method's accuracy of plus/minus one week (see Methods), demonstrating the possibility to obtain accurate real-time estimates of downtime. 
The error is larger in Juchit\'an de Zaragoza because of the larger fluctuations of the estimates due to a lower sample of businesses. 

\begin{figure*}[h]
    \centering
    \includegraphics[width=\textwidth]{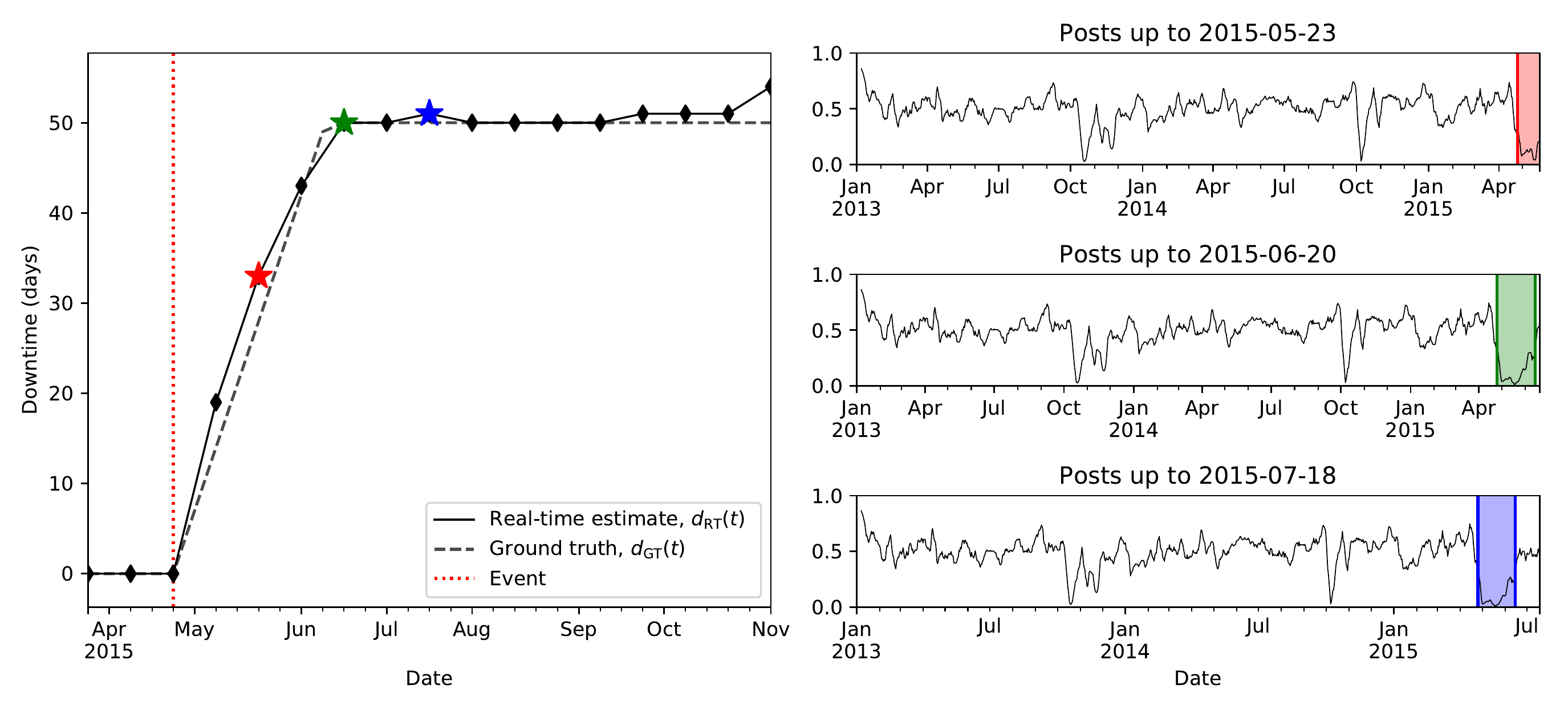}
    \caption{Event detection shown in real-time. Data was cropped at different end dates (up to seven months after the event), and the process was applied to this data. Shown are the time series for one, two and three months after the event in Kathmandu respectively. The black dashed line indicates the perfect real-time detection for the downtime predicted using all of the data.}
    \label{process_real}
\end{figure*}

\section{Discussion}
A framework has been proposed to determine the 
recovery status of small businesses in the aftermath of 
natural disasters through the analysis of the businesses' posting activity on their public pages on the social media site Facebook.

This framework relies on the assumption that businesses tend to publish more posts when they are open and fewer when they are closed, hence analysing the aggregated posting activity of a group of businesses over time it is possible to infer when they are open or closed. 
This assumption is verified in three case studies, where we successfully estimate the downtime of businesses in three urban areas affected by 
different natural disasters:
the 2015 Gorkha earthquake in Nepal, the 2017 Chiapas earthquake in Mexico and the 2017 hurricane Maria in Puerto Rico. 
We validate our predictions using various data sources: field interviews, official tourism data, Facebook surveys and other studies available in literature.

The methodology proposed has the potential to be highly informative to predict the return to activity of businesses in a specific area with an accuracy of plus/minus one week, provided that there is a sufficient use of Facebook by the businesses in the area. 
Given the widespread diffusion of social media websites, the method can be easily applied to multiple regions at the same time, enabling to nowcast the recovery status of areas affected by natural and artificial disasters on a global scale. 
The proposed methodology for the detection of anomalous events in non-stationary time series is of general applicability to all cases where the main signal is composed by the aggregation of a large number of individual components, for example phone users' calling activities or internet users' visits to web pages.

\section{Materials and Methods}
\subsection{Natural disasters}
We report a short description for each natural disaster: 

\subsubsection*{2015 Gorkha Earthquake (25-04-2015; Kathmandu, Nepal).}
According to worldwide news results, the 2015 Gorkha Earthquake was one of the worst natural disasters in Nepalese history, killing around 9,000 people, and injuring nearly 22,000 people. It occurred on April 25th, with a magnitude of Mw 7.8. Its epicentre was located 77km North West of Kathmandu (east of the Gorkha district). Hundreds of aftershocks were reported following the earthquake, five of which registered Mw above 6. The most notable aftershock occurred on the 12th of May, killing a further 200 people and injuring a further 2,500. 8.1 million Nepalese citizens were thought to be affected by the earthquake \cite{OfficeoftheResidentCoordinator2015}. The International Monetary Fund  \cite{IMF} label Nepal as an emerging and developing country. The event affected significantly the all country with detected damage in the city of Kathmandu especially in the North West area of the city \cite{Goda_2015}.

\subsubsection{Chiapas Earthquake (07-09-2017; Chiapas, Mexico).}
The 2017 Mw 8.2 Chiapas Earthquake was the second strongest earthquake recorded in Mexico's history (and most intense globally in 2017), triggered in the Gulf of Tehuantepec, just off the south coast of Mexico. Multiple buildings in the city closest to the epicentre, Juchitán de Zaragoza, were reduced to rubble with people sleeping outdoors due to fears of further damage \cite{ACAPS2017}. The main hospital in Juchitán also collapsed causing severe disruption to medical services in the area, with 90 reported dead in Mexico. Food shops were also affected, with prices rises due to closures, and fears from looting causing more closures. According to local authorities, roughly 30\% of houses were damaged in the earthquake. This is probably aggregated by the lack of earthquake resilient houses, predominantly made of adobe block with concrete bond beam and mud wall constructions \cite{UNITAR-UNOSAT2017}. \$150 million dollars was transferred to the National Disaster Fund (FONDEN) to aid health, education and roads infrastructure/reconstruction efforts \cite{UNICEF2017}.

\subsubsection{Hurricane Maria (20-09-2017; San Juan, Puerto Rico).}
Puerto Rico, along with Florida and the US Virgin Islands, was hit by a Category 5 hurricane on September 20th, 2017 causing significant damage to infrastructure \cite{UnitedStatesDepartmentofEnergy2017}, affecting 100\% of the population \cite{FEMA2018}, with an estimated 45\% of islanders being without power for three months after the event \cite{Shermeyer2018}. Hurricane Maria is estimated to have caused \$94 billion in damage, with as many as 60,000 homes still lacking roofs as of December 2016 \cite{ReliefWebPue}. As of August 2018, FEMA have approved 50,650 loans to small businesses proving \$1.7 billion to aid in recovery, in total obliging \$21.8 billion in relief efforts \cite{FEMA2018}.

\subsection{Data collection}
\label{sec:datacoll}

On the social media website Facebook, businesses can set up their own pages (called \say{walls}) to advertise and interact with users, with approximately 65 million local businesses pages created as of 2017 \cite{facebook_numbers}. These business pages include the latitude and longitude of the business, allowing for the possibility of spatial analysis on the posts created in a given region. 

Publicly available business data has been retrieved from Facebook via the Facebook Graph API, in areas affected by natural disasters;  
we found 
11,818 businesses that published 1.1m posts in Kathmandu, NP, 
10,894 businesses that published 2.2m posts in San Juan, PR and
1,728 businesses that published 62k posts
in Juchitán de Zaragoza, MX. 
The locations of the businesses in the three regions considered are shown in Figure~\ref{fig:all}a-c. 
It should be noted that a visible post on the wall of a business can be made by the owner of the business, or a member of the public. Facebook differentiates these messages by offering API access to the wall's posts (messages made by the owner) or the wall's feed (messages made by everyone, including the author). For the purpose of this study, we collect only the messages made by the owner of the page.

Once businesses' pages had been collected from Facebook, an additional set of queries were ran using the API to collect the time-stamps in which each business made a post. The contents of the posts were not stored.

\subsection{Data processing}
\label{sec:dataproc}

We define the aggregated time series describing the behaviour of the entire region, $r(t)$, as the 
time series of the total number of posts made by all the businesses, $B$:

\begin{equation}
r(t) = \sum_{i \in B}{x_i}(t)
\end{equation}

where $x_i(t)$ is the number of posts made by business $i$ on day $t$. %
The raw time series $r(t)$ display a reduced activity during a period of time following the disasters, as shown in Figure~\ref{fig:all}d-f. However, it is difficult to estimate the significance and the length of the downtime from the raw time series because it is non-stationary. 
In particular, there is an increasing trend in the average number of posts, due to widespread adoption to Facebook by local businesses over time. 
Additionally, the time series present various seasonality effects, such as reduced posting rates during holidays like Christmas and New Year in Puerto Rico, which can be more or less pronounced on different years. 
Some of these recurring patterns are quasi-periodic, for example Dashain Festival in Nepal does not occur on the same day every year, and this makes it harder to remove them. 
Moreover, businesses have very different posting rates, with some being up to two orders of magnitude more active than the average. Such very active businesses can introduce significant biases and lead to over or under estimate the activity level of the entire region, when the overall number of businesses is small. This is the case of the activity spikes in Juchit\'an de Zaragoza in 2015, which are caused by a very small number of businesses.  

To account for these issues, we develop a method to transform the raw time series into a detrended and rescaled series that allows us to clearly identify the periods of anomalous activity and measure their length. 
The methodology to process the time series is composed of four steps: 
(1) Single business Probability Integral Transformation (PIT) and aggregation, (2) Shift and rescale, (3) Mean and variance correction, (4) Aggregated Probability Integral Transformation. 

\subsection*{Step 1: Single business PIT and aggregation.}
In step (1) the raw time series of the posts of each business is transformed into the time series of the corresponding \say{mid-quantiles}. Formally, this is obtained using a procedure similar to the Probability Integral Transform (PIT)  \cite{Angus_1994}. Let $x_i(t)$ be the number of posts of business $i$ on a given day $t$ and let $P_{X_i}(x)$ be the empirical Cumulative Distribution Function (CDF) denoting the fraction of days of the current year when business $i$ made less than $x$ posts. 
We define the corresponding mid-quantile for $x_i(t)$ as 
$q_i(t) = [ P_{X_i}(x_i(t)) + P_{X_i}(x_i(t) + 1) ] / 2$. 
We use the CDF of the current year, instead of considering the business' entire lifetime, to take into account long-term changes in a business' posting behaviour. Using the mid-quantile variable $q$ instead of the raw number of posts $x$ helps to reduce the bias due to outliers, for example days with an unusually high number of posts, and the bias caused by businesses with mean posting rates much higher than the average.
Aggregating the transformed data, we have the time series 
\begin{equation} \label{eq:pit}
r_{PIT}(t) = \sum_{i \in B} q_i(t) \, .
\end{equation}

\begin{figure*}[h]
    \centering
    \includegraphics[width=\textwidth]{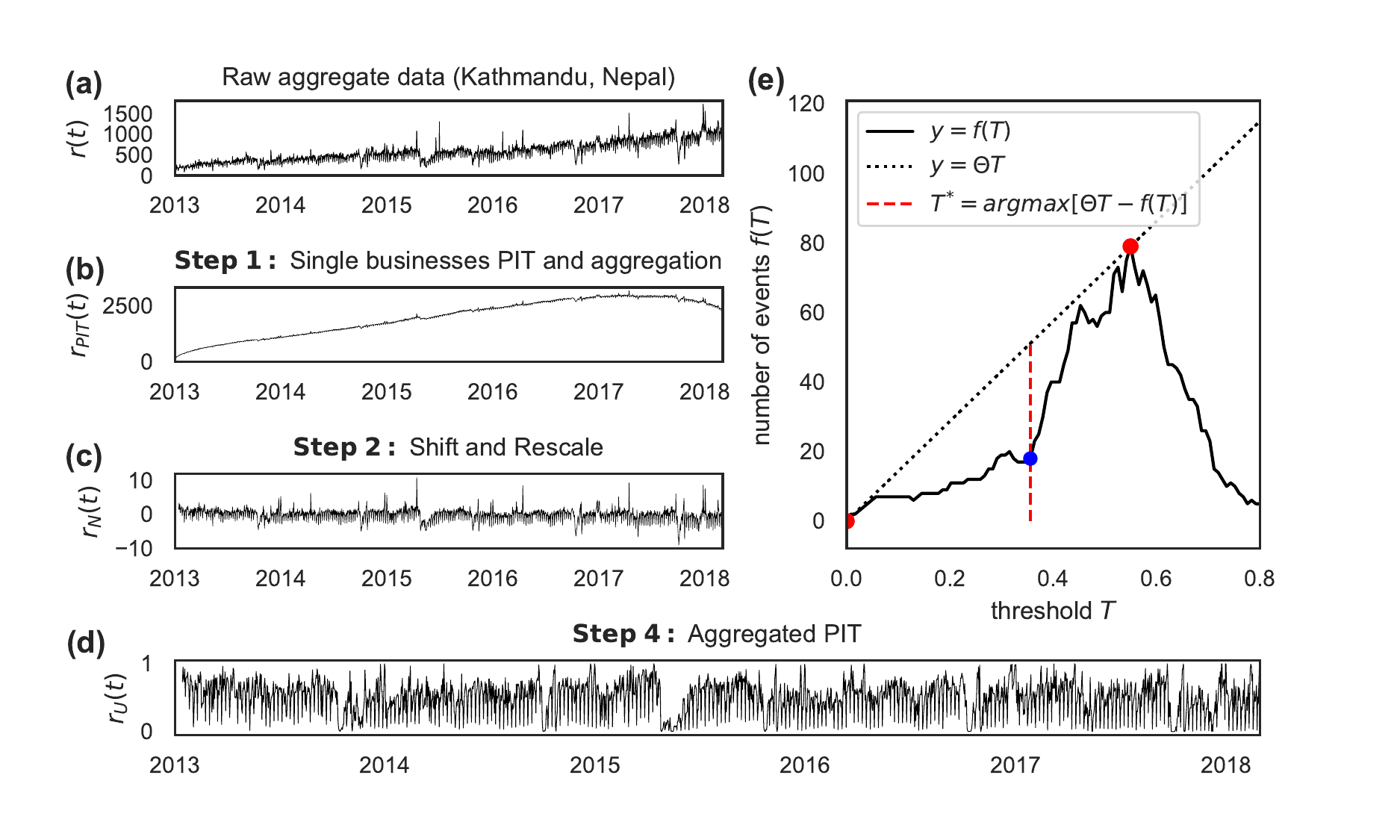}
    \caption{
    Steps of the process to estimate the downtime of small businesses applied to Kathmandu, Nepal. \textbf{(a)} Raw data \textbf{(b)} PIT transformed data \textbf{(c)} Shifted and rescaled data \textbf{(d)} Probability Integral Transform applied on the aggregated and transformed time series \textbf{(e)} Threshold and event detection. }
    \label{process}
\end{figure*}

\subsection{Step 2: Shift and rescale.}
After the PIT, the mid-quantiles $q_i(t)$ are expected to be uniformly distributed between 0 and 1. Hence, the aggregated transformed data $r_{PIT}(t)$ of Equation~\ref{eq:pit} is expected to follow the Irwin-Hall distribution, which is the distribution of the sum of $n(t)$ uncorrelated Uniform random variables between 0 and 1. Here $n(t)$ is the total number of businesses with a Facebook account on day $t$. When $n(t)$ is large, the Central Limit Theorem ensures that the distribution of $r_{PIT}(t)$ is well approximated by a Normal distribution with mean $n(t) / 2$ and variance $n (t) / 12$. Hence, by appropriately shifting and rescaling each day of $r_{PIT}(t)$ we define the time series 
\begin{equation} \label{eq:normal}
r_{N}(t) = (r_{PIT}(t) - n(t) /2) / \sqrt{n(t) / 12} \, ,
\end{equation}
whose distribution is a Standard Normal with mean 0 and variance 1 for all days $t$. To perform the transformation of Equation~\ref{eq:normal} we have to estimate $n(t)$. 
The number of businesses with an active Facebook account on a given day %
is estimated by recording the first and last post date for each business $i$, denoted as $t^{(f)}_i$ and $t^{(l)}_i$ respectively. We make the assumption that a business remains active between these dates, plus a period of two more days to be sure that it has closed. This period was chosen to give an unbiased estimate of the number of active businesses at the tail end of the data collected. 
Formally, we estimate the number of businesses with an active Facebook account on day $t$ as 
$
n(t) =  | \{ i :  t^{(f)}_i \leq t \leq t^{(l)}_i+2  \} |
$. 
The time series $r_N(t)$ for Kathmandu is shown in Figure~\ref{process}c and those for the other regions considered are shown in the Supplementary Materials. 

\subsection{Step 3: Mean and variance correction.} 
The assumption that the businesses' posting activities are independent and the estimate of the number of businesses on Facebook $n(t)$ are reasonable, but may not be perfectly exact. In fact, a small correlation between businesses' posting activity may be present even during normal (non-emergency and non-holiday) conditions and this would affect the variance of $r_{PIT}$. On the other hand, deviations in the estimate of $n(t)$ could also change the mean of $r_{PIT}$. To correct for these possible sources of bias, we fit and remove a linear trend from $r_{N}(t)$, so that its mean becomes zero, and we divide it by its standard deviation, so that its variance becomes one. The resulting corrected time series, $\tilde{r}_{N}(t)$, for the three regions are shown in the Supplementary Materials.

\subsection{Step 4: Aggregated PIT.} 
Finally, we apply the Probability Integral Transform  to the aggregated time series of normally distributed variables, $\tilde{r}_{N}(t)$, to obtain a time series of variables uniformly distributed between 0 and 1. 
This is done using the CDF of the Standard Normal distribution, $P_N$: 
\begin{equation} \label{eq:unif}
r_U(t) = P_N(\tilde{r}_N(t)) \, .
\end{equation}
The final time series $r_U(t)$ for Kathmandu is shown in Figure~\ref{process}d and those for all the regions are shown in Figure~\ref{fig:final_all}. 
The four steps of the data processing for all the regions considered are shown in the Supplementary Materials. 

\subsection{Downtime detection}

\label{sec:downtime}

The length of downtime is estimated as follows. 

We consider the rolling weekly mean of the transformed time series $r_U(t)$. This is primarily to account for differing posting behaviour over the course of a week (e.g. weekdays vs. weekends). 
We define an {\it event} as a period of time when the weekly mean of $r_U(t)$ is below a given threshold value, $T$, for more than 7 consecutive days. 
The value of the threshold is set using a method aimed at detecting all periods when posting activity is significantly below the characteristic range of normal activity fluctuations. 
This value is found using the following \say{elbow method}. 
First, the number of potential downtime events are recorded at multiple threshold values; the number of events for a given threshold value is denoted as $f(T)$ and it is shown in Figure~\ref{process}e for Kathmandu. 

Second, we identify the threshold value, $T^{\ast}$, at which the number of events begins to rapidly rise. This point marks the transition between the size of activity fluctuations due to anomalous events and the  characteristic size of normal activity fluctuations. 
We define $T^{\ast}$ as the abscissa of the point on the \say{elbow} of function $f(T)$, i.e. the value of $T$ that maximises the distance between $f(T)$ and the line connecting the origin to a point on $f$ such that its slope, $\theta$, is maximum (see Figure~\ref{process}e): 
$T^{\ast} = \argmax_{T} (\theta T - f(T) )$. 
The dashed vertical line in Figure~\ref{process}e denotes the value of $T^{\ast}$ for Kathmandu. The events detected in the three regions using this method are shown in Figure~\ref{fig:final_all}.

\subsection{Validation}
\textbf{Research field data.}
Surveys from four areas in Kathmandu were used to verify the results. These surveys were part of a research field mission in 2016 to examine the extent of the earthquake damage, with regions chosen that vary in population density, construction methods and traffic concentration \cite{DeLuca_2018}.  These surveys record the length of downtime for each business, along with the reason for closure. Regions A and B both report damage as the main reason for business closure. Regions C and D both report scare as reason for closure in the region, it was reported that shop owners feared another aftershock and remained closed until the risk of danger was reduced. 

\textbf{Tourism data.} Historical tourism data has been retrieved from the Puerto Rico Tourism Company~\cite{puerto_tourism}, listing the cruise passenger movement in port in Old San Juan. We compute the Year over Year (YoY) change, preserving seasonal trends, calculating the percentage change of each month to the same month in the previous year.

\textbf{Facebook surveys.} To collect validation for the region collected in Mexico, we sent surveys to the recorded businesses asking the question \say{Were you affected by the disaster, and if so, for how long were you closed?}. We sent this survey to 52 of the businesses found via Facebook in Juchit\'an de Zaragoza, with a response rate of 30\% (16 responses).

\section{Data availability}
The raw data that supports the findings of this study is not available due to the Facebook Platform Policy. Derived (anonymised) data that supports the findings of this study are available from the authors upon reasonable request.

\section{Author contributions}
All authors contributed to the conception and design of the study. FDL conducted the field interviews, RE collected the social media data and analysed the results. All authors wrote and reviewed the manuscript.

\section{Competing interests}
The authors declare no competing interests.

\section{Acknowledgements}
RE is grateful for the support received by EPSRC and the 2017/18 Cabot Institute Innovation Fund. 
FDL acknowledges the support of the Leverhulme Trust (RPG2017-006, GENESIS project). 
FS is supported by EPSRC First Grant EP/P012906/1.

\bibliographystyle{ieeetr}
\bibliography{bib.bib}

\end{document}


\maketitle

\section{Data collection}

As Facebook's search does not return every business from a given location and search radius, we divided each region into 250m sections and pass over these sections when searching (instead of searching the whole region). Searches are made at radii 250m, 375m and 500m to ensure as many businesses are returned as possible. Duplicate results were then subsequently removed.

\section{Survey locations}

\begin{figure*}[]
\begin{framed}
    \centering
    \includegraphics[width=\textwidth]{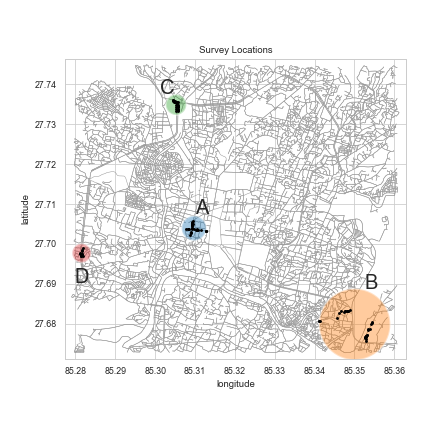}
    \caption{Survey locations in the study by De Luca et al. \cite{DeLuca_2018}}
    \label{fig:surv_regions}
\end{framed}
\end{figure*}

We highlight the locations in which surveys were taken in Kathmandu, Nepal. 92 surveys were taken in total. To validate our results, we also ran the described process on each area, finding all businesses in the smallest circle to contain all surveys from the centre of the survey region (Figure \ref{fig:surv_regions}). We find 532 businesses in region A, 163 businesses in region B, 24 businesses in region C, and 73 businesses business in region D. 
We report the average downtime in each region in Table \ref{tab:surv_regions}. 
 
\begin{table}[h!]
\centering
\begin{tabular}{p{2.75cm}p{3.5cm}p{3.5cm}p{3.5cm}}
\toprule
{} &  Mean downtime &  Standard deviation &   Number of surveys \\
\midrule
Area A           &           29 &          44 &                        27 \\
Area B           &           15 &          20 &                        31 \\
Area C           &           61 &          88 &                        25 \\
Area D           &           78 &          62 &                        10 \\
\bottomrule
All regions &           41 &          62 &                        94 \\
\bottomrule
\end{tabular}
\caption{
Downtime (days) reported in four areas in Kathmandu, Nepal, from the surveys taken by De Luca et al. \cite{DeLuca_2018}}
\label{tab:surv_regions}
\end{table}

\section{Process for each region}

\begin{figure*}[]
\begin{framed}
    \centering
    \includegraphics[width=\textwidth]{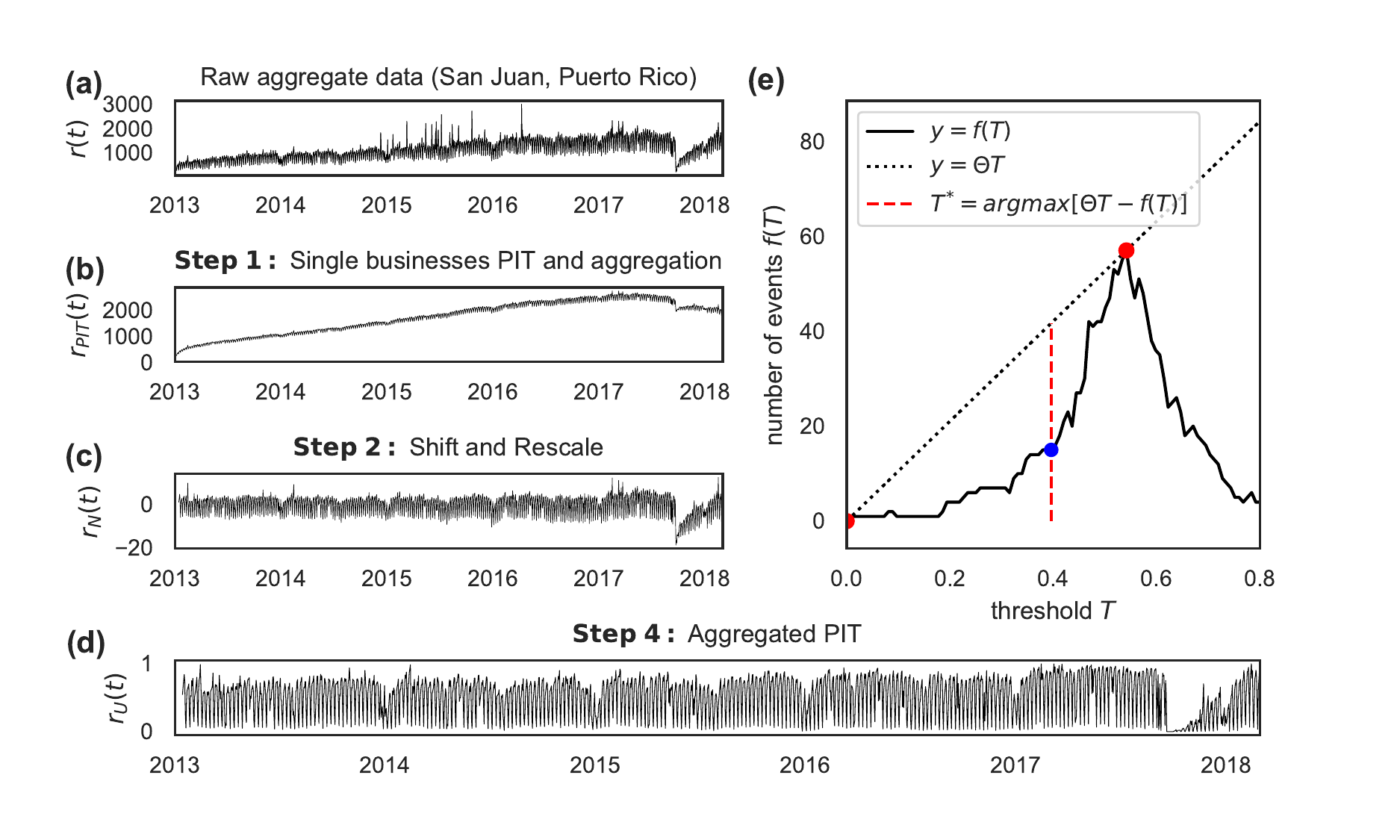}
    \caption{Steps of the process applied to San Juan. \textbf{(a)} Raw data \textbf{(b)} PIT transformed data \textbf{(c)} Shifted and rescaled data \textbf{(d-e)} Threshold and event detection.}
    \label{fig:proc}
\end{framed}
\end{figure*}

\begin{figure*}[]
\begin{framed}
    \centering
    \includegraphics[width=\textwidth]{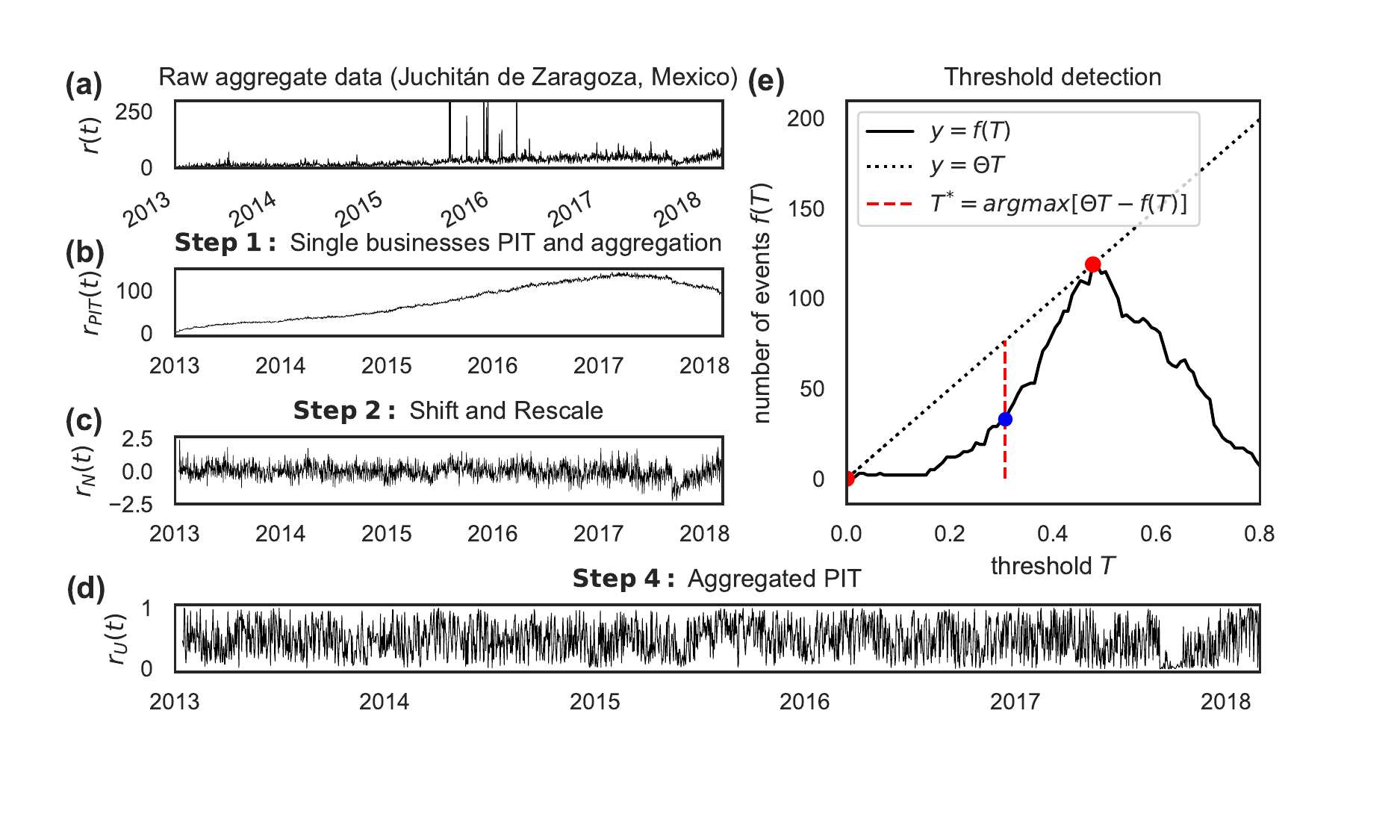}
    \caption{Steps of the process applied to Juchit\'an de Zaragoza. \textbf{(a)} Raw data \textbf{(b)} PIT transformed data \textbf{(c)} Shifted and rescaled data \textbf{(d-e)} Threshold and event detection.}
    \label{fig:proc2}
\end{framed}
\end{figure*}

We show the framework applied to the two natural disasters not shown in the main text (Figure \ref{fig:proc}, Figure \ref{fig:proc2}).

\section{Variance correction}

We make the assumption of independence between businesses' posting habits, and also the assumption that businesses are always active between their first and last post date. Whilst this is reasonable, it does not always hold - for example correlation between businesses’ posting activity may be present during normal (non-emergency and non-holiday) conditions, affecting the variance of $r_{PIT}$. Also deviations in the estimate of n(t), where we have not perfectly predicted the number of active businesses could also change the mean of $r_{PIT}$. To correct for these possible sources of bias, we fit and remove a linear trend from $r_N(t)$, so that its mean becomes zero,  and we divide it by its standard deviation,  so that its variance becomes one.

\section{Choice of business filter}

\begin{figure*}[]
\begin{framed}
    \centering
    \includegraphics[width=\textwidth]{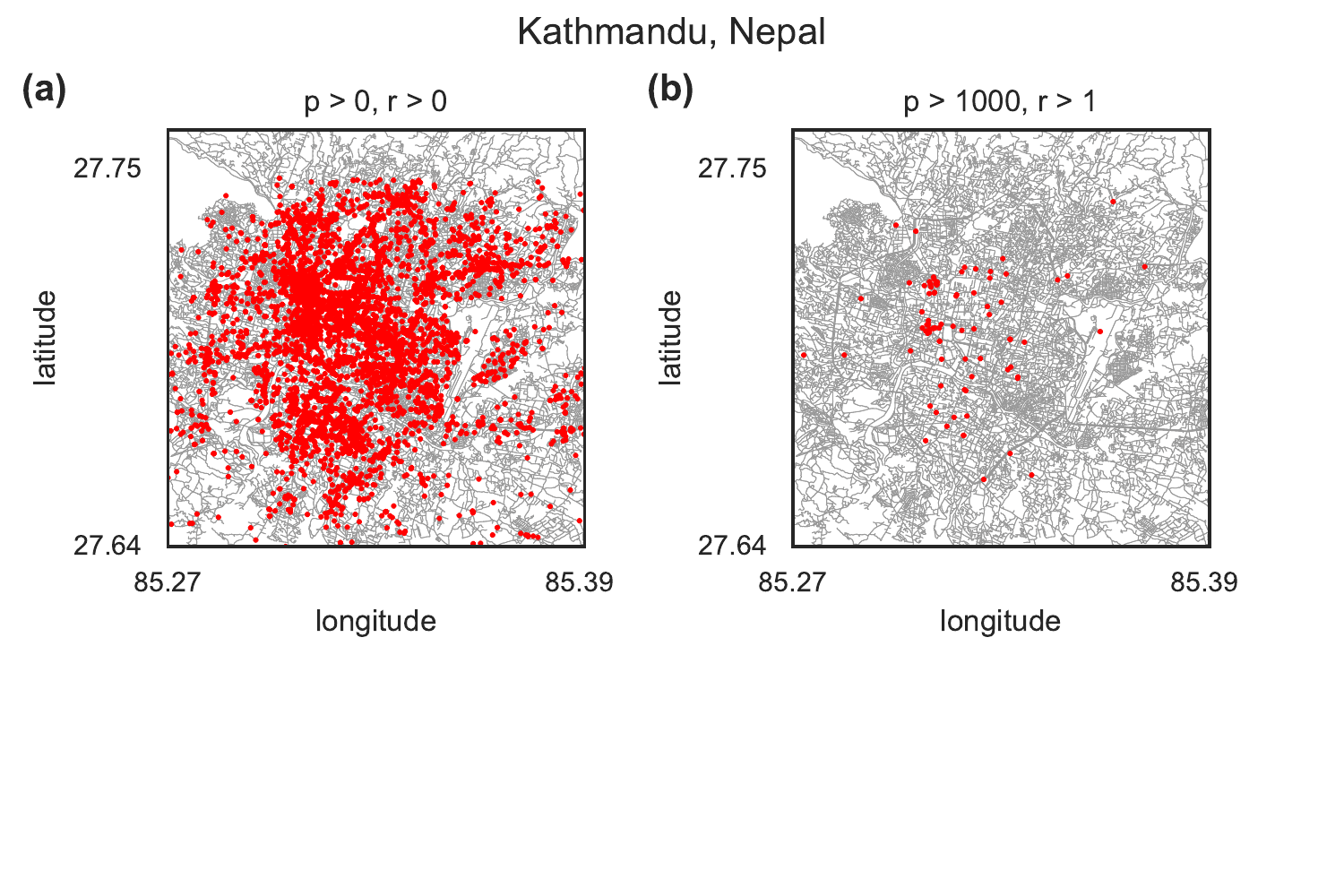}
    \vspace{-100pt}
    \caption{Imposing a high threshold on the quality of the business (at least 1000 posts, at least 1 post per day), the majority of the businesses found are in the centre of Kathmandu, meaning that the final estimate is not representative of the whole region.}
    \label{fig:quality}
\end{framed}
\end{figure*}

The reported downtime for each region was calculated using all of the collected businesses that had posted at least once. We can filter businesses by the rate at which they post and the total number of posts they have made to determine how dependent the process is on business \say{quality}. In the case of Kathmandu, our predictions stay relatively consistent (Table \ref{tab:kat_downtimes}), with less downtime being reported as we introduce higher filters - the majority of businesses that meet the criteria in these cases are located in the most densely commercially populated area in the centre of the city (Figure \ref{fig:quality}).

\section{Business sampling}

\begin{figure*}
\begin{framed}
    \centering
    \includegraphics[width=0.49\textwidth]{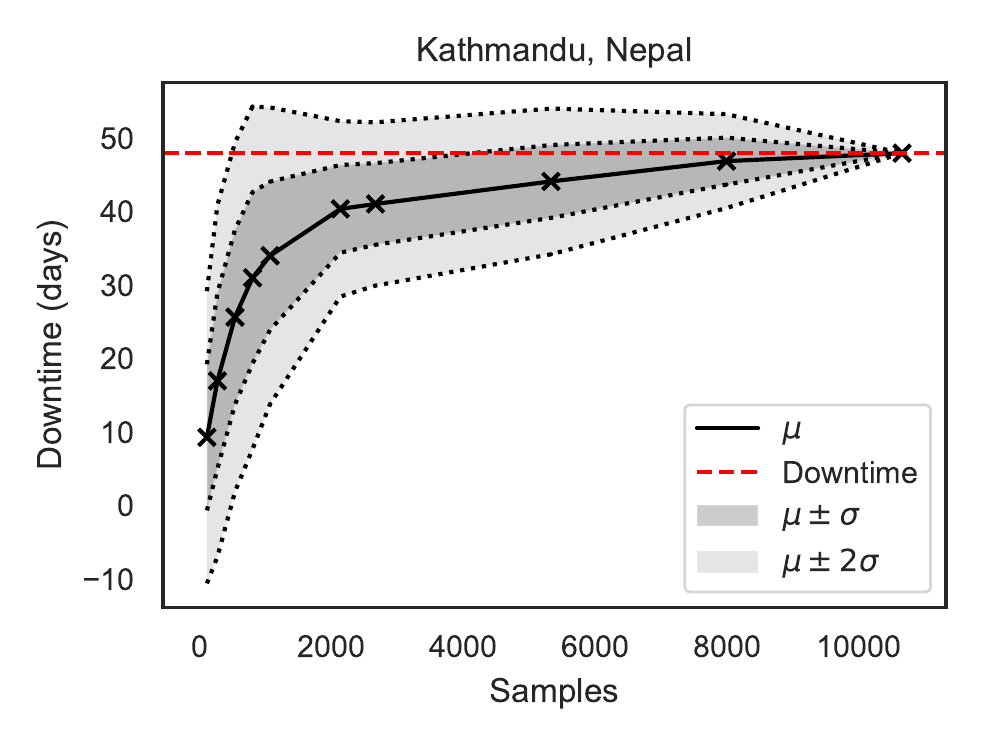}
    \includegraphics[width=0.49\textwidth]{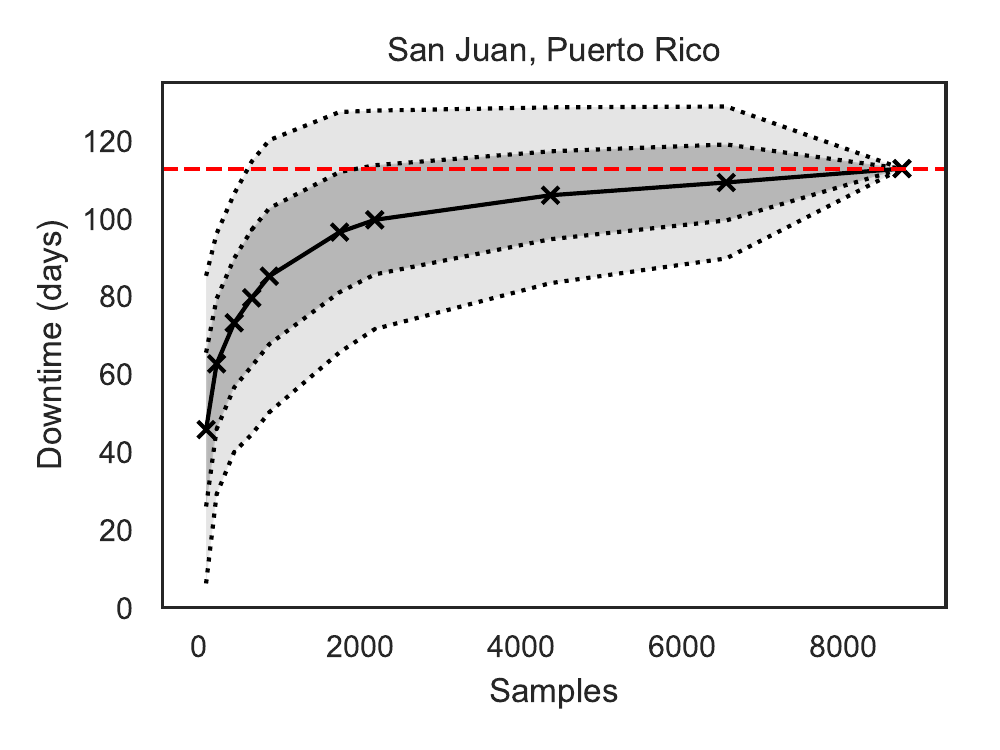}
    \caption{Business samples of different sizes are taken in Kathmandu, Nepal (left) and San Juan, Puerto Rico (right). Downtimes are calculated, and then this process is repeated. Average downtimes shown with two standard deviations from the mean. }
    \label{fig:samples}
\end{framed}
\end{figure*}

We highlight the impact that under-sampling a region can have on the final prediction for the entire region. For each sample size, we draw 1000 realisations (with replacement) and computer the average downtime predicted using our framework. In Kathmandu we are able to see the downtime for the whole region (within one standard deviation from the mean) within 3000 businesses. In San Juan we are able to see the downtime for the whole region (within one standard deviation from the mean) within 2000 businesses (Figure \ref{fig:samples}).

\section{Validation, San Juan }

To estimate the true downtime in San Juan, we look at the Year over Year reported incoming tourists via the port in Old San Juan \cite{puerto_tourism}. 

\section{Downtime detection}

The start date, end date and the length of downtime of all the events detected in the three regions analysed are reported in Table~\ref{tab:kathmandu_events} (Kathmandu, Nepal), Table~\ref{tab:puerto_events} (San Juan, Puerto Rico), Table~\ref{tab:chiapas_events} (Juchitán de Zaragoza, Mexico). 

\section{Real time detection}

\begin{figure*}
\begin{framed}
    \centering
    \includegraphics[width=\textwidth]{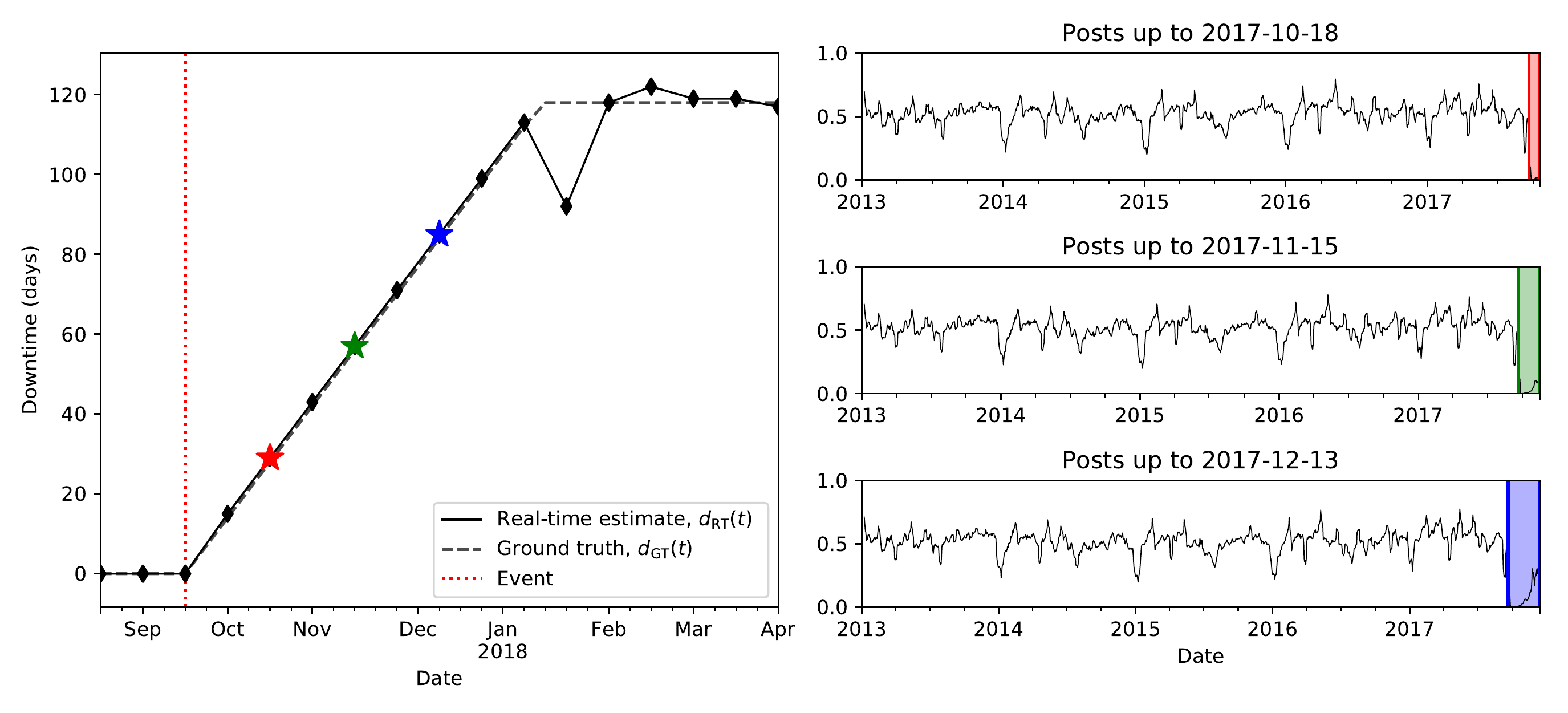}
    \caption{Event detection shown in real-time. Data was cropped at different end dates (up to seven months after the event), and the process was applied to this data. Shown are the time series for one, two and three months after the event in San Juan respectively. }
    \label{fig:realtime_pr}
\end{framed}
\end{figure*}

\begin{figure*}
\begin{framed}
    \centering
    \includegraphics[width=\textwidth]{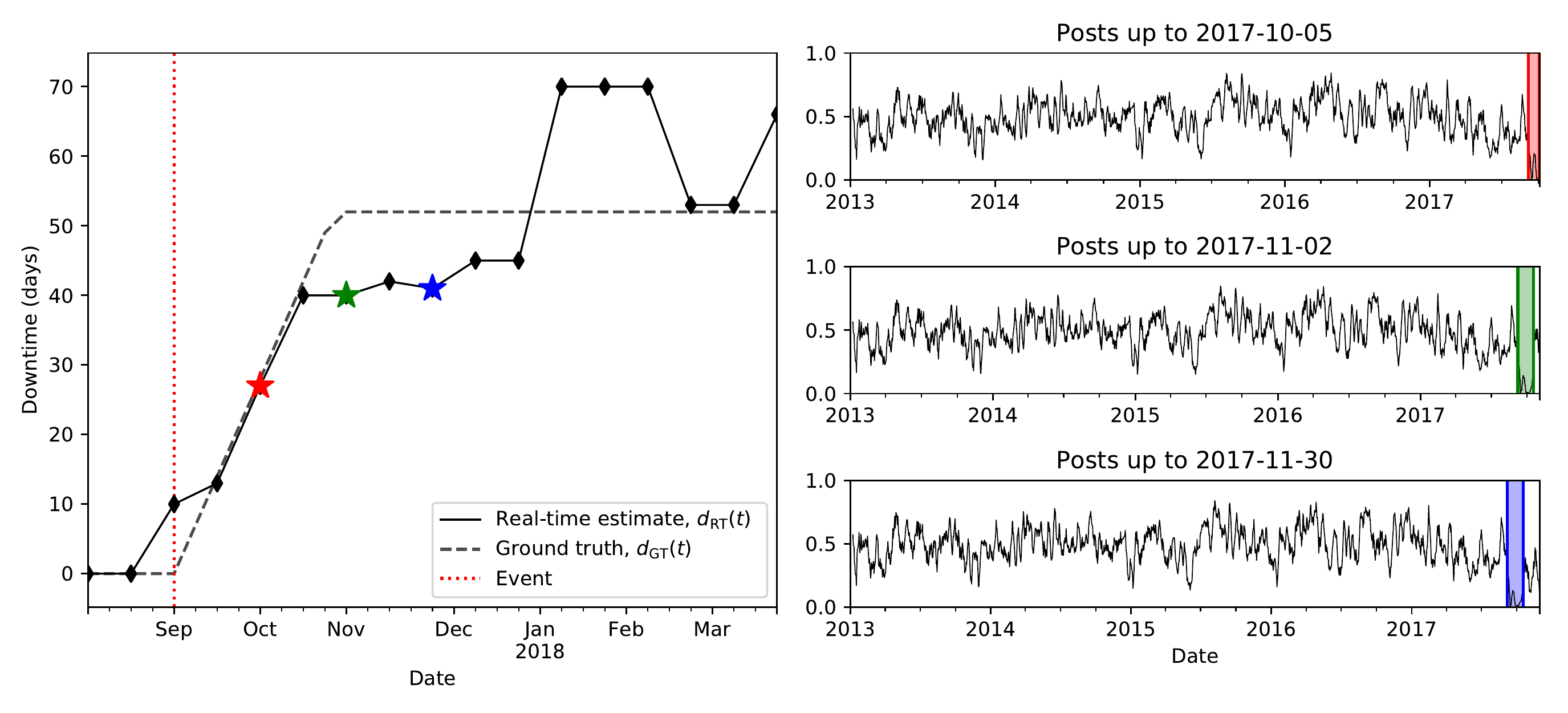}
    \caption{Event detection shown in real-time. Data was cropped at different end dates (up to seven months after the event), and the process was applied to this data. Shown are the time series for one, two and three months after the event in Juchit\'an de Zaragoza respectively. }
    \label{fig:realtime_c}
\end{framed}
\end{figure*}

We show the downtime predicted for different cut-off dates in the regions of San Juan (Figure \ref{fig:realtime_pr}) and Juchit/'an de Zaragoza (Figure \ref{fig:realtime_c}). 

\begin{table*}[h!]
    \centering

\begin{tabular}{llrl}
\toprule
 Start date &    End date &  Duration (days) & Event \\
\midrule
 2013-10-13 &  2013-10-31 &               19 &    Dashain 2013 (5th Oct - 17th Oct)   \\
 2013-11-06 &  2013-11-27 &               22 &    Tihar 2013 (1st Nov - 5th Nov)   \\
 2014-01-07 &  2014-01-20 &               14 &   Unknown    \\
 2014-10-03 &  2014-10-14 &               12 &     Dashain 2014 (24th Sep - 7th Oct) \\
 2015-04-27 &  2015-06-15 &               50 &     \textbf{Gorkha Earthquake (25th April)}  \\
 2015-10-21 &  2015-11-02 &               13 &     Dashain 2015 (13th Oct - 26th Oct) \\
 2016-10-10 &  2016-10-25 &               16 &     Dashain 2016 (1st Oct - 15th Oct) \\
 2017-05-04 &  2017-05-21 &               18 &    Unknown   \\
 2017-09-28 &  2017-10-14 &               17 &     Dashain 2017 (20th Sep - 4th Oct) \\
 2017-11-26 &  2017-12-18 &               23 &    Unknown   \\
 2018-02-04 &  2018-02-11 &                8 &    Unknown   \\
\bottomrule
\end{tabular}

    \caption{Detected events in Kathmandu, Nepal}
    \label{tab:kathmandu_events}
\end{table*}

\begin{table*}[]
    \centering

\begin{tabular}{llrl}
\toprule
 Start date &    End date &  Duration (days) &                 Event \\
\midrule
 2014-12-28 &  2015-01-12 &               16 &        Christmas / New Year 2015 \\
 2015-12-29 &  2016-01-11 &               14 &        Christmas / New Year 2016 \\
 2016-12-30 &  2017-01-09 &               11 &        Christmas / New Year 2017 \\
 2017-09-21 &  2018-01-16 &              118 &  \bf{Hurricane Maria (20th Sep)} \\
\bottomrule
\end{tabular}
    
    \caption{Detected events in San Juan, Puerto Rico. }

    \label{tab:puerto_events}
\end{table*}

\begin{table*}[]
    \centering

\begin{tabular}{llrl}
\toprule
 Start date &    End date &  Duration (days) & Event \\
\midrule
 2015-05-26 &  2015-06-09 &               15 &    Unknown   \\
 2016-11-01 &  2016-11-09 &                9 &    All Saints' Day / Day of the Dead (1/2 Nov)   \\
 2017-09-10 &  2017-10-31 &               52 &   \textbf{M8.2 Chiapas Earthquake (7th Sep)}    \\
 2017-11-02 &  2017-11-13 &            12 &    All Saints' Day / Day of the Dead (1/2 Nov)   \\
 2017-11-23 &  2017-11-30 &                8 &   Revolution Day (20th Nov)    \\
\bottomrule
\end{tabular}

    \caption{Detected events in Juchit\'an de Zaragoza, Mexico.}
    \label{tab:chiapas_events}
\end{table*}

\bibliographystyle{ieeetr}
\bibliography{biblio.bib}


\begin{figure*}[]
\begin{framed}
    \centering
    \includegraphics[width=\textwidth]{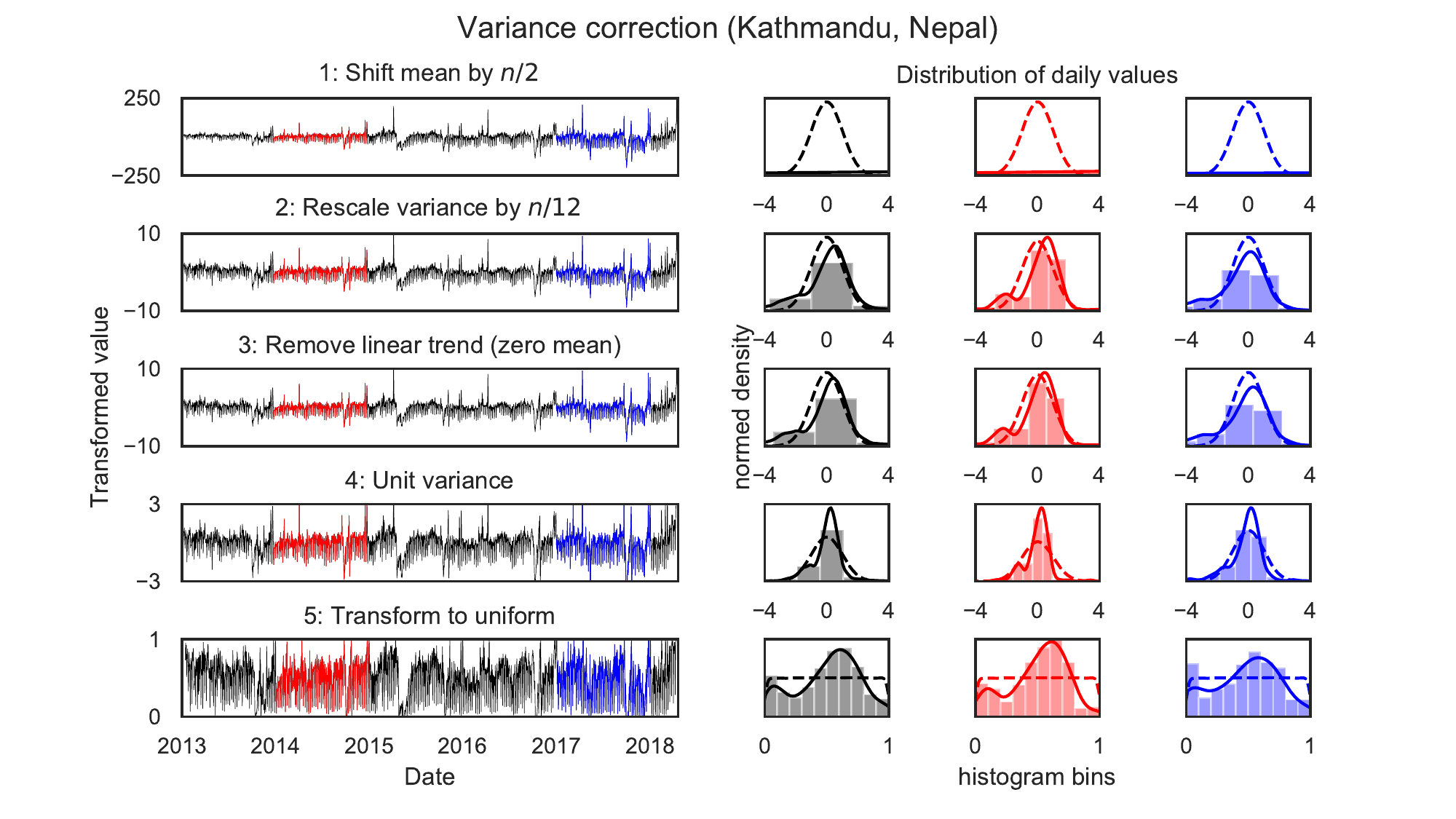}
    \caption{Variance correction applied to $r_N(t)$ in Kathmandu, Nepal.}
    \label{fig:var0}
\end{framed}
\end{figure*}

\begin{figure*}[]
\begin{framed}
    \centering
    \includegraphics[width=\textwidth]{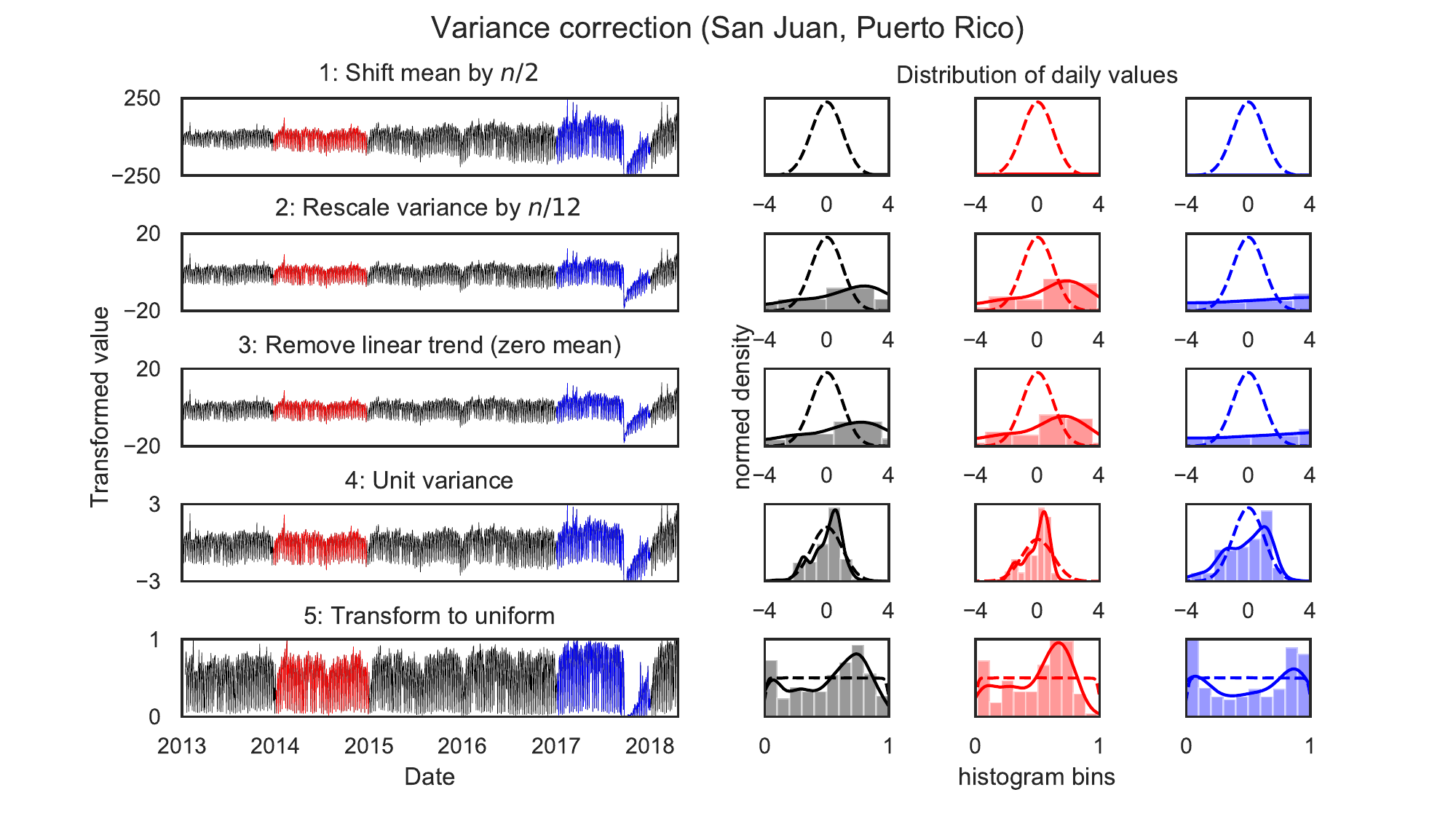}
    \caption{Variance correction applied to $r_N(t)$ in San Juan, Puerto Rico}
    \label{fig:var1}
\end{framed}
\end{figure*}

\begin{figure*}[]
\begin{framed}
    \centering
    \includegraphics[width=\textwidth]{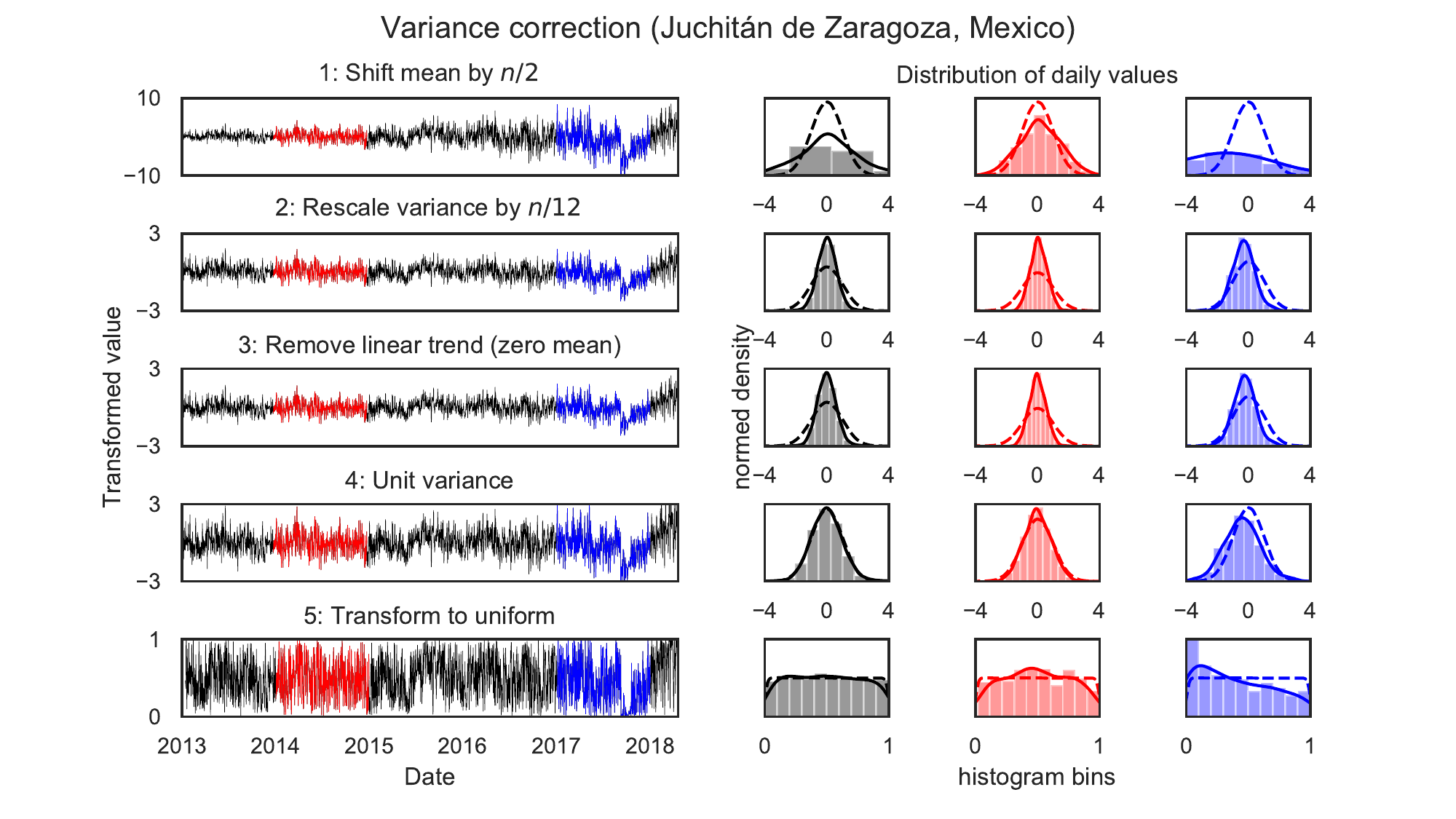}
    \caption{Variance correction applied to $r_N(t)$ in Juchit\'an de Zaragoza, Mexico.}
    \label{fig:var2}
\end{framed}
\end{figure*}

\begin{table*}[]
    \centering

\begin{tabular}{LLLLLL}
\toprule
{_{\text{\# posts}} / ^{\text{Daily post rate}}} &      0 &     1/7 &     1/2 &     5/7 &     1 \\
\midrule
1   &  50_{(10386)} &  48_{(4080)} &  47_{(1625)} &  34_{(1216)} &  37_{(900)} \\
100 &   48_{(2523)} &  47_{(1758)} &   35_{(526)} &   35_{(297)} &  51_{(161)} \\
200 &   47_{(1391)} &  47_{(1288)} &   36_{(448)} &   34_{(259)} &  51_{(147)} \\
300 &    47_{(952)} &   47_{(952)} &   47_{(398)} &   34_{(236)} &  51_{(138)} \\
400 &    46_{(686)} &   46_{(686)} &   35_{(363)} &   34_{(220)} &  37_{(130)} \\
\bottomrule
\end{tabular}
    \caption{
    Overall downtime reported over the whole of Kathmandu, Nepal, with different filters on the businesses (by the number of posts they have made, and the daily average posting rate). Number of businesses that meet the criterion are listed in brackets for each reported downtime. }
    \label{tab:kat_downtimes}
\end{table*}

\begin{table*}
\centering
\begin{tabular}{LLLLLLL}
\toprule
{_{\text{\# posts}} / ^{\text{Daily post rate}}} &      0 &     1/7 &     1/2 &     5/7 &     1 \\
\midrule
1   &  118_{(8459)} &  120_{(4833)} &  113_{(2276)} &  112_{(1649)} &  111_{(1140)} \\
100 &  120_{(3462)} &  120_{(2916)} &  113_{(1292)} &   112_{(824)} &   111_{(485)} \\
200 &  117_{(2412)} &  117_{(2314)} &  117_{(1183)} &   112_{(765)} &   111_{(457)} \\
300 &  117_{(1840)} &  117_{(1840)} &  116_{(1088)} &   112_{(724)} &   111_{(436)} \\
400 &  117_{(1513)} &  117_{(1513)} &  113_{(1021)} &   112_{(691)} &   111_{(422)} \\
\bottomrule
\end{tabular}
    \caption{
    Overall downtime reported over the whole of San Juan, Puerto Rico with different filters on the businesses (by the number of posts they have made, and the daily average posting rate). Number of businesses that meet the criterion are listed in brackets for each reported downtime. }
    \label{tab:pr_downtimes}
\end{table*}

\begin{table*}[]
    \centering

\begin{tabular}{LLLLLLL}
\toprule
{_{\text{\# posts}} / ^{\text{Daily post rate}}} &      0 &     1/7 &     1/2 &     5/7 &     1 \\
\midrule
1   &  52_{(552)} &  48_{(287)} &  46_{(139)} &  57_{(104)} &  46_{(84)} \\
100 &  48_{(128)} &  48_{(105)} &   48_{(34)} &   48_{(24)} &  43_{(15)} \\
200 &   39_{(73)} &   40_{(71)} &   44_{(30)} &   46_{(21)} &  43_{(13)} \\
300 &   45_{(49)} &   45_{(49)} &   46_{(26)} &   45_{(18)} &  80_{(12)} \\
400 &   10_{(34)} &   10_{(34)} &   10_{(23)} &   47_{(16)} &  43_{(11)} \\
\bottomrule
\end{tabular}
    \caption{
    Overall downtime reported over the whole of Juchit\'an de Zaragoza, Mexico, with different filters on the businesses (by the number of posts they have made, and the daily average posting rate). Number of businesses that meet the criterion are listed in brackets for each reported downtime. }
    \label{tab:chi_downtimes}
\end{table*}

\begin{figure}[h!]
    \centering
    \includegraphics[width=0.5\textwidth]{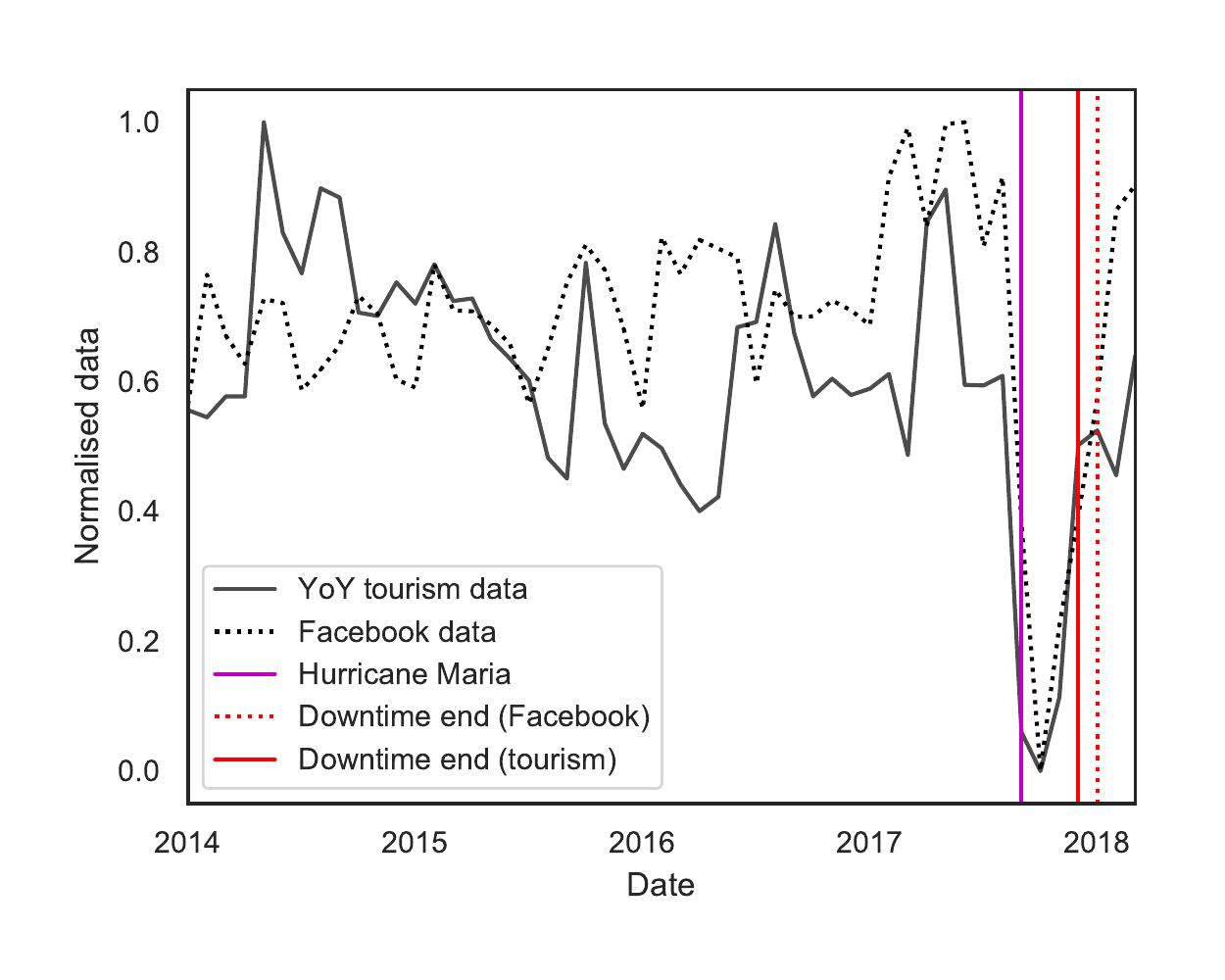}
    \caption{
    Normalised Year-Over-Year tourism data in San Juan (solid) vs normalised signal data from our framework (dotted).}
    \label{fig:tourism}
\end{figure}

